# Increasing the Number of Underrepresented Minorities in Astronomy Through K-12 Education and Public Outreach (Paper II)

*An Astro2010 State of the Profession Position Paper*
March 2009

**Authored by**: The AAS Committee on the Status of Minorities in Astronomy (CSMA), **with endorsement from:**
National Society of Hispanics in Physics - David J. Ernst, Pres. (Vanderbilt University), Marcel Agueros (Columbia University), Scott F. Anderson (University of Washington), Andrew Baker (Rutgers University), Adam Burgasser, (Massachusetts Institute of Technology), Kelle Cruz (Caltech), Eric Gawiser (Rutgers University), Anita Krishnamurthi (University of Maryland, College Park), Hyun-chul Lee (Washington State University), Kenneth Mighell (NOAO), Charles McGruder (Western Kentucky University), Dara Norman (NOAO), Philip J. Sakimoto (University of Notre Dame), Kartik Sheth (Spitzer Science Center), Dave Soderblom (STScI), Michael Strauss (Princeton University), Donald Walter (South Carolina State University), Andrew West (MIT)
UW Pre-Map staff - Eric Agol (Faculty Project Leader), Jeremiah Murphy, Sarah Garner, Jill Bellovary, Sarah Schmidt, Nick Cowan, Stephanie Gogarten, Adrienne Stilp, Charlotte Christensen, Eric Hilton, Daryl Haggard, Sarah Loebman Phil Rosenfield, Ferah Munshi (University of Washington)

Primary Contact
Dara Norman
NOAO
950 N. Cherry Ave
Tucson, AZ 85719
dnorman@noao.edu,
520-318-8361

*Abstract*
In order to attract, recruit and retain underrepresented minority students to pursue Astronomy and related fields, we must ensure that there continues to be a well qualified pool of graduate and undergraduate students from which to recruit. This required pool of people are today's elementary, middle and high school students. The Astronomy community must be proactive in demonstrating the importance of pursing scientific study and careers to these students and their parents. Only by actively engaging these communities can U.S Astronomy hope to increase the numbers of minority PhDs and continue to be a leader in Astronomical discovery and knowledge.

1. **Statement of the problem:**

Increasingly the challenge to our nation is the decrease in the highly skilled, technical workforce. U.S. students are scoring lower on tests in science and math than their counterparts worldwide. The American education system is failing to produce the technically literate workforce needed to advance astronomy research. For every 100 US students entering $9^{th}$ grade, only 18% will graduate within the 7-10 years with a college associates or bachelors degree. Furthermore, only 30% of US college students major in science fields (Hunt, 2006). The changing demographics of the U.S. over the next 10+ years will exacerbate this deficit of technically skilled workers. The underrepresented minority workforce (adults age 25-64) is growing the fastest, but these are the same workers who are most likely to be under-educated for positions in the fields of science and technology (see Figures 1 & 2). If these education gaps continue (and continue to grow), we face a situation where there could be a sharp decrease in the working population who have the basic qualifications needed to pursue higher education in science fields (see Figure 3).

In order to keep our technological workforce strong, we must be proactive in increasing the opportunities for minority students to succeed in the areas of math and science. Until we increase the numbers of minority students coming out of our K-12 school system who are proficient in math and science, we will not be able to involve these students in the astronomy enterprise[1].

Formal astronomy classes are rarely offered to K-12 students until perhaps their last one or two years of high school, however astronomy remains an important gateway science because of its ability to capture the imaginations of people of all ages (Ward et al, 2007, AANM, 2001). The public popularity of astronomy can and should be used to improve science literacy, including for those from underrepresented minorities and underserved areas[2].

---

[1] By "Astronomy Enterprise" we mean Astronomy and related fields that include Physics, engineering and computer science. See the associated Positional Paper. "Increasing the Number of Underrepresented Minorities in Astronomy at the Undergraduate, Graduate, and Postdoctoral Levels," hereafter, Paper I.

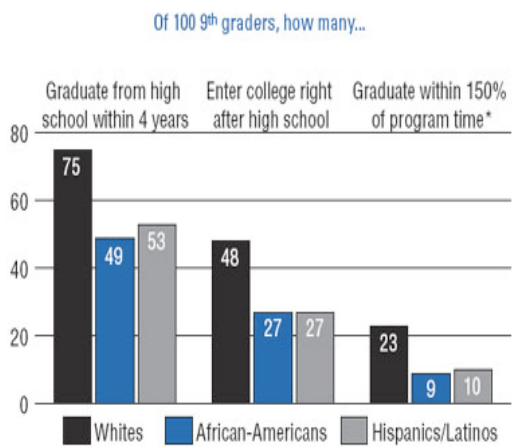

Figure 1. The U.S. educational pipeline by race/ethnicity. These are the numbers for 2001. This data does not account for transfer students. Note also that statistics from a 2004 paper published by the Urban Institute show that in 2001 high school graduation rates in the US were 51% for American Indians. Source: Analysis by NCHEMS (www.higheredinfo.org) based on data from NCES common Core Data; IPEDS2002 Fall Enrollment Survey; IPEDS 2002 Graduation Rate Survey.

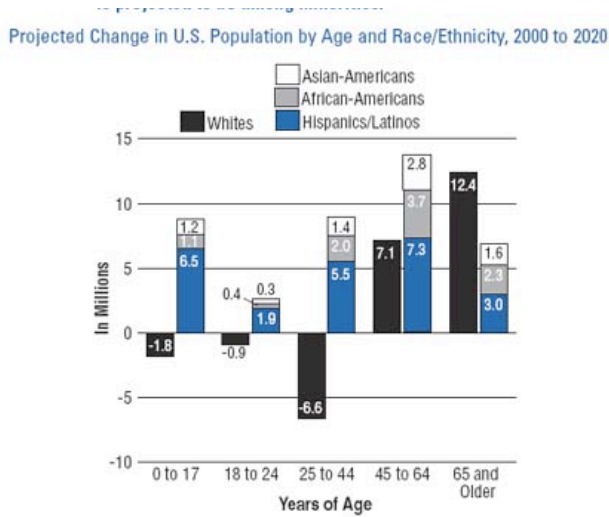

Figure 2: The greatest portion of the U.S population growth from ages 0 to 44 is projected to be among minorities. Source: U.S. Census Bureau, 5% Public Use Microdata Samples [based on the 2000 Census].

The previous astronomical decadal review, "Astronomy and Astrophysics in the New Millennium" (2001), presents recommendations for increasing the role of astronomy in K-12 education and public outreach. Here we build on these recommendations by identifying steps that need to be taken to expand K-12 Astronomy education to underserved communities, specifically to underrepresented minority students.

## 2. **Solutions:**

All students, including those with underrepresented minority backgrounds, need to be engaged in science inquiry at early stages in their schooling careers with knowledgeable and experienced science teachers[3]. This engagement must be made in both formal (Cuevas et al., 2005) and informal (Bell et al., 2009) settings where creative learning and thinking can occur. Furthermore, this engagement must involve long-term exposure to science inquiry and scientific ways of thinking (Jolly et al., 2004).

---

[2] Here we used the term 'underserved' regions or areas to refer to poor urban and rural locations. While these may not be exclusively predominantly minority areas, often they are.

[3] We cite here the policy recommendations made in the National Science Teachers Association document, "Rising Above the Storm: Science Education in the 21st Century". (www.nsta.org/about/olpa/risingabove.aspx)

Astronomers must be proactive in promoting science and math interest at the K-12 levels by interacting with students, teachers and the public in underserved communities in order to nurture an understanding of the field and its benefits to the broader community (AANM 2001).

**The Astronomer's role in engaging minority K-12 students**

Minority students should be exposed early to science learning and activities. Early on students are interested in almost everything, but as they progress to middle school and later, programs may be needed to engage or re-engage students in science disciplines (Clewell & Campbell, 2000, Jolly, Campbell and Perlman 2004).

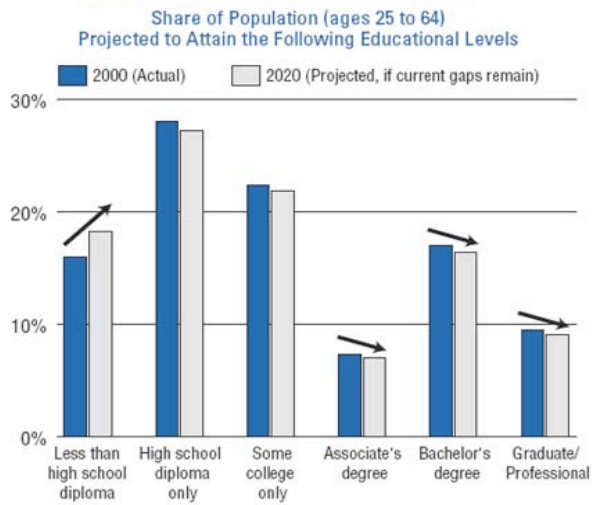

Figure 3: Educational gaps. If current educational gaps remain, there will likely be a substantial increase in the percentage of workforce with less than a high school diploma and therefore declines in the higher levels of education [based on the 2000 Census].

Astronomy is a good introductory science topic for younger students because physical science concepts are readily accessible to students through simple activities (Ward, R.B., et al., 2007). For example, very young children are able to recognize the moon and know that sometimes it is up during the day as well as at night and that it does not always appear to be the same shape. With some help, younger children can even recognize that at a given time of day, the moon will be in a different part of the sky on consecutive nights. In other words, even kindergarten students can make astronomical observations. These types of observations require a few clear nights/days, but they don't require equipment or even a dark sky. Students in the most heavily light polluted inner city can participate in this kind of activity. This is the power of Astronomy to introduce students (even very young students) to science inquiry.

Programs like the Astronomical Society of the Pacific's Project Astro and Family Astro engage young students in the experience of science inquiry[4]. These programs have proven to be popular with teachers, students and the Astronomers who partner with them. They have been successful, not only in teaching students how to engage in scientific thinking through astronomy examples, but also in providing role models for children who may not know any 'real' scientists. Project Astro has made some inroads into a few underserved areas through an NSF grant to expand the

---

[4] Websites for Project Astro, www.astrosociety.org/education/astro/project_astro.html and Family Astro , www.astrosociety.org/education/family.html .

initiative. With this money the project was able to give priority to proposed sites that would prioritize outreach to such areas (private communication). However, to reach a critical mass of minority students, programs like this must be expanded to include more of these communities. One way to do this is to expand funding for programs that partner Astronomers and teachers to other educational institutions, not just schools, in minority serving areas. For example, community-based after school programs should be included in funding programs. These facilities, engage the same youths and families year after year thus providing the continuity needed to engage and retain student interest in science

Informal education can be just as influential to improving the interest and learning of students in the area of science (Bell, 2009, Yager & Falk, 2008). Science centers, museums and astronomy camps are important to advancing the science learning and understanding environment for K-12 students. For example, astronomy camps[5] where students participate in astronomical observations with telescopes at dark sites provide experiences similar to those of real scientists and allow students access to peer groups with similar interests. These types of interactions are important since they allow students the freedom to engage in activities supported by a community. The International Year of Astronomy presents an opportunity to begin these kinds of informal learning experiences for minority students, but these initiatives must be continued and funded over longer periods of time (more than one year) if they are to have an impact on broadening participation in Astronomy (e.g. Sakimoto et al. 2009).

**The Astronomer's role in engaging teachers in underserved areas**
The research literature on science education and curriculum development is robust and mature[6]. Astronomers should horizontally[7] partner with professional organizations of teachers, science centers [8] and local teachers to effectively use this information and not "reinvent the wheel". A number of astronomy centers and departments support teacher professional development workshops[9] where inquiry based methods of astronomy instruction that comply with state or federal standards are taught. As with all teachers, those in underserved areas must be prepcleware to be effective instructors for science classes and those who need additional help must have access to efficacious training programs[10]. Professional development workshops for teachers must be funded and expanded in ways that more actively include teachers in underserved and minority areas if we are to increase participation of students from these areas. As is the case with several universities currently, astronomers should continue to be involved teacher enrichment programs as they are expanded. The AAS should take an active role in helping to evaluate and keep track of these programs so that they are more easily visible to interested teachers.

---

[5] For example see, http://www.astronomycamp.org/

[6] Examples of materials include a. LHS Great Explorations in Math and Science, Lawrence Hall of Science, University of Science Berkeley, 2002.
b. Loucks-Horsley, S., Hewson, P.W., Love, N. and Stiles, K., Designing Professional Development for Teachers of Science and Mathematics, The National Institute for Science Education Corwin Press, INC. Thousand Oaks, CA 1998.

[7] By 'horizontal partnering' we mean, as in the Paper 2, genuine partnerships in which each party is responsible for 'mission critical' parts of the larger project. That is, where the Astronomer-teacher partnerships are mutually respectful, collaborative and sustaining.

[8] For example such organizations include, the American Association of Physics Teachers (AAPT), the Association of Science and Technology Centers (ASTC), American Association of Museums

[9] Examples of University based teacher enhancement programs include: a. Beck-Winchatz, B. and Barge, J. 2003 "A New Graduate Space Science Course for Urban Elementary and Middle School Teachers at DePaul University in Chicago" The Astronomy Education Review, Issue 1, Volume 2:120-128 and b. McDonald Observatory and University of Texas at Austin's Summer Workshops at the Observatory, see mcdonaldobservatory.org/teachers/profdev/

**The Astronomer's role in engaging the underserved public**
Smith & Hausafus (1998) show that students maintain higher scores in math and science if parents help them to see the importance of these subjects in their future careers. This is particularly true for ethnic minority students (see Sakimoto et al., 2007). Therefore, it is important that astronomers commit resources and time to engagement in local underserved communities to encourage citizen (parental) interest and understanding of scientific results in the field of astronomy. Many astronomy departments already engage in public outreach programs like public lecture series, department open houses and star parties. It is important that these activities actively be expanded to include more minority participation and access. This can be achieved through expanded and sustained advertisement of outreach events and the hosting of activities to include minority accessible locations like African-American and Latino themed museums, Minority Serving Institutions, e.g. local HBCUs, HSIs, Tribal colleges[11], community centers and organizations as well as churches. In order to sustain this engagement in the minority community, astronomy departments need to be flexible in supporting and aiding students, postdocs and faculty to participate in these activities, perhaps even sponsoring TA positions for interested students. These science education and outreach activities are important to the health and success of the astronomical enterprise and must be recognized as such by the astronomical community.

3. <u>**Specific Recommendations:**</u>

**Recommendations for colleges, universities, & national centers**
1. Commit resources (money, interested staff, time) to support engagement in local underserved communities, to encourage citizen interest and understanding of scientific results and the field of astronomy.
2. Form 'horizontal' two-way partnerships with teacher's professional organizations and local teachers to determine effective ways to engage minority K-12 students and/or their teachers.

---

[10] From recommendations made in the National Science Teachers Association document, "Rising Above the Storm: Science Education in the 21st Century". (www.nsta.org/about/olpa/risingabove.aspx)
[11] See Paper 2 appendix for maps of HBCUs, HSIs and Tribal colleges.

3. Form partnerships with minority serving institutions —both academic and community based— to identify the best ways to interact with these communities
4. Provide support and aid for students, postdocs and faculty to participate in outreach activities, perhaps even sponsoring TA positions for interested students.
5. Start or support, by involvement in, local academic and mentoring programs for minority students that bridge the transition from high school to college[12].

**Recommendations for funding agencies**
1. Provide opportunities for continuity of funding for astronomy education programs. The most successful programs that engage minority students and teachers in underserved areas can achieve continuity by establishing Federal funding cycles in the same way that research is funded. The current practice of awarding seed funding with restrictions against coming back for more is deleterious to the success of such programs.
2. Continue to fund and expand astronomy education programs that focus both on student enrichment and teacher professional development in underserved communities.
3. Fund efforts to expand informal learning in underserved communities.

**Recommendations for professional societies**
1. Create and maintain current networks and databases for teachers/students with information on available astronomical programs and workshops.
2. Play a lead role in identifying exceptionally effective K-12 outreach and education programs and work to see that they are adopted widely, particularly in underserved communities.
3. Endorse efforts to institute inquiry methods of teaching astronomy wherever possible, in classroom & informal teaching, as well as in citizen science forums.

**References:**
Bell, P., Lewenstein, B., Shouse, A.W. and Feder, M.A.(Eds), 2009, "Learning Science in Informal Environments: People, Places and Pursuits",. Washington, DC: National Academies Press.

Cuevas, P., Lee, O., Hart, J. and Deaktor, D. 2005, "Improving Science Inquiry with Elementary Students of Diverse Backgrounds", Journal of Research in Science Teaching, 42 (3), 337.

---

[12] For examples, see programs like Pre-Map detailed in the appendix of Paper I. A few other programs that support transitions for minority students from high school to college are the Posse Foundation (www.possefoundation.org) and MIT's Project Interphase (//web.mit.edu/ome/programs-services/interphase/index.html).

Clewell & Campbell, 2002, "Taking Stock: Where we've been, where we are going", Journal of Women and Minorities in Science and Engineering, vol. 8, pp. 255–284.

Hunt Jr, J. and Tierney, T. 2006. "American Higher Education: How Does It Measure Up for the 21[th] Century?" San Jose, CA: National Center for Public Policy and Higher Education.

Jolly, E., Campbell, P.B., Perlman, L. 2004). "Engagement, Capacity and Continuity: A Trilogy for Student Success", GE Foundation ([www.smm.org/papers](www.smm.org/papers))

National Center on Education and the Economy 2007, "Tough Choices or Tough Times: The report of the new commission on the Skills of the American Workforce"

National Center for Public Policy and Higher Education Policy Alert, Income of U.S. Workforce Projected to Decline if Education Doesn't Improve. ([www.highereducation.org/reports/pa_decline/index.shtml](www.highereducation.org/reports/pa_decline/index.shtml))

National Research Council (2001) "Astronomy and Astrophysics in the New Millennium" Washington, DC: National Academy Press. (AANM 2001)

Sakimoto, P., Luckey, V., Landsberg, R.H, Hawkins, L., and Porro,I. 'Preparing for the 2009 International Year of Astronomy' ASP Conference Series, Vol. 400, c2008 M. G. Gibbs, J. Barnes, J. G. Manning, and B. Partridge, eds.

Smith, F.M., and Hausafus, C.O. 1998, "Relationship of Family Support and Ethnic Minority Students' Achievement in Science and Mathematics", Science Education 82 (1) 111

Ward, R.B., Sadler, P.M., and Shapiro, I.I., 2007 "Learning Physical Science through Astronomy Activities: A Comparison between Constructivist and Traditional Approaches in Grades 3-6" Astronomy Education Review, Vol. 6, No. 2, pp. 1–19,

Yager, R.E. and Falk, J. (Eds) 2008 "Exemplary Science in Informal Education Settings", Arlington, VA.: NSTA press